\documentclass[aps, prd, twocolumn, nofootinbib, floatfix, superscriptaddress]{revtex4-2}
\usepackage{subfigure}
\usepackage{epsfig}
\usepackage{amsmath}
\usepackage{amsfonts}
\usepackage{amssymb}
\usepackage{amssymb,amsmath,amsfonts}
\usepackage{xfrac}
\usepackage{color}
\usepackage[utf8]{inputenc}
\usepackage{graphicx}
\usepackage{dcolumn}
\usepackage{bm}
\usepackage{tikz}

\usepackage{tcolorbox}
\usepackage{xurl}
\usepackage{hyperref}
\hypersetup{colorlinks, citecolor=red, linkcolor=bluscuro, urlcolor=bluscuro}
\definecolor{rossos}{cmyk}{0,1,1,0.55}
\definecolor{bluscuro}{rgb}{0.15, 0.2, .85}
\definecolor{bluchiaro}{cmyk}{1,.3,0.,0.1}

\begin{document}

\preprint{APS/123-QED}

\title{Dark energy from topology change induced by microscopic Gauss-Bonnet 
wormholes}

\author{Stylianos A. Tsilioukas}
\email{tsilioukas@sch.gr}
\affiliation{\mbox{Department of Physics, University of Thessaly, 35100 Lamia, Greece}}
\affiliation{\mbox{National Observatory of Athens, Lofos Nymfon, 11852 Athens, Greece}}
 
\author{Emmanuel N. Saridakis}
\email{msaridak@noa.gr}
\affiliation{\mbox{National Observatory of Athens, Lofos Nymfon, 11852 Athens, Greece}}
\affiliation{CAS Key Laboratory for Researches in Galaxies and Cosmology, 
Department of Astronomy, \\
University of Science and Technology of China, Hefei, 
Anhui 230026, P.R. China}
\affiliation{\mbox{Departamento de Matem\'{a}ticas, Universidad Cat\'{o}lica del 
Norte, 
Avda.
Angamos 0610, Casilla 1280 Antofagasta, Chile}}

\author{Charalampos Tzerefos}
\email{chtzeref@phys.uoa.gr}
\affiliation{\mbox{National Observatory of Athens, Lofos Nymfon, 11852 Athens, Greece}}

\affiliation{\mbox{Department of Physics, National \& Kapodistrian University of Athens, Zografou Campus GR 157 73, Athens, Greece}}

\begin{abstract}
 
It is known that the appearance of microscopic objects with distinct topologies 
and different Euler characteristics, such as instatons and wormholes, at the 
spacetime-foam level in Euclidean quantum gravity approaches leads to spacetime 
topology changes. Such changes, in principle, may affect the field equations 
that arise through the semiclassical variation procedure of gravitational 
actions. Although in the case of Einstein-Hilbert action the presence of 
microscopic wormholes does not lead to any non-trivial result, when the 
Gauss-Bonnet term is added in the gravitational action, the above effective 
topological variation procedure induces an effective cosmological constant that 
depends on the Gauss-Bonnet coupling and the wormhole density. Since the latter 
in a dynamical spacetime is in general time-dependent, one obtains an effective 
dark energy sector of topological origin.

\end{abstract}

\maketitle

\section{\label{Introduction}Introduction}

According to overwhelming observations of various origins, the Universe entered 
the phase of accelerated expansion in the recent cosmological past 
\cite{SupernovaSearchTeam:1998fmf,SupernovaCosmologyProject:1998vns,
WMAP:2003elm,SDSS:2003eyi,Allen:2004cd,SNLS:2005qlf}.
The simplest explanation is the introduction of a positive  cosmological 
constant $\Lambda$, nevertheless such a consideration faces    
the ``cosmological constant problem'', since quantum field 
theoretical analysis predicts a value up to $120$ orders of magnitude larger than the 
observed one   \cite{Weinberg:1988cp,Sahni:1999gb}. Additionally, the resulting 
cosmological concordance model, namely $\Lambda$CDM   paradigm, seems to 
exhibit possible tensions at the 
phenomenological level, such as the $H_0$ \cite{DiValentino:2020zio} and 
$\sigma_8$ tensions \cite{DiValentino:2020vvd} (for a review see 
\cite{Abdalla:2022yfr}).

In order to alleviate the aforementioned issues one can follow two main 
directions. The first is to  consider a dynamical cosmological constant, or 
more 
generally the concept of dark energy  \cite{Copeland:2006wr,Cai:2009zp}, in the 
framework of general relativity. The second direction is to modify  the 
underlying gravitational theory by constructing novel theories with richer 
behavior at cosmological scales 
\cite{CANTATA:2021ktz,DeFelice:2010aj,Capozziello:2011et,Cai:2015emx,
Nojiri:2017ncd}. Moreover, one could incorporate  more radical 
considerations in order to explain dark energy, such as   the framework of 
holographic dark energy \cite{Li:2004rb,Saridakis:2020zol,Drepanou:2021jiv}.
 
Modified gravity, apart from alleviating the cosmological issues, has the 
additional advantage of improved quantum behavior \cite{AlvesBatista:2023wqm}, 
since general relativity is non-renormalizable  \cite{Addazi:2021xuf}.
In particular, when higher-order curvature terms are included in the 
Einstein-Hilbert Langrangian, they tend to eliminate the divergences 
\cite{Stelle:1976gc}, which led to an increased interest in the 
construction of higher order theories of gravity 
\cite{Lanczos:1938sf,Lovelock:1971yv,Starobinsky:1980te,Capozziello:2002rd}. In 
these lines, in heterotic string theory the  Gauss-Bonnet (GB) invariant is 
included in the Langrangian due to its role in regulating   divergences 
\cite{Stelle:1976gc,Zwiebach:1985uq,Boulware:1985wk}. Furthermore,
from all the higher order terms, the  GB one has extensive 
and crucial implications, since it is the Euler density in four dimensions 
(4D),  and thus according to the Chern-Gauss-Bonnet Theorem \cite{Chern:1945} it 
is a topological invariant in 4D, while it preserves the local supersymmetry of 
the heterotic string \cite{Gross:1986mw}. Finally, in M-theory the contribution 
of the GB term is essential in canceling the divergences that appear in the 
beta-function at high energies, thus facilitating the renormalization of the theory 
\cite{Vafa:1994tf}.

On the other hand, it is known that  in the framework of  Euclidean quantum 
gravity there are solutions such as instatons and
wormholes, which exhibit different topology from the background 
\cite{Gibbons:2011dh}. Wormholes have also been investigated in astrophysics \citep{DeFalco:2020afv,DeFalco:2021klh,DeFalco:2021ksd,DeFalco:2021btn,DeFalco:2023kqy,DeFalco:2023twb}.  If one assumes that these objects appear at 
the spacetime-foam level \cite{HAWKING1978349}, then the spacetime background 
becomes topologically  dynamical. The effects of topology changes have been thoroughly 
studied in the literature.   For instance, in \cite{Geroch:1967fs} it was 
shown that topology change in classical Lorenztian spacetime manifolds leads to the development of singularities, and based on this result,  Anderson and 
DeWitt  argued that quantum field theory is inconsistent with 
such singularities \cite{Anderson:1986ww}. Additionally, in the framework of 
quantum gravity the feasibility of topology change has been supported by 
several investigations  
\cite{Horowitz:1990qb,Borde:1994tx,Dowker:2002hm,Csizmadia:2009dm}. In 
\cite{Sorkin:1997gi} Sorkin argued that topology change is required in order 
for quantum gravity to be consistent, while    in \cite{Gibbons:2011dh} Gibbons
showed that in the context of Euclidean Quantum Gravity  the Wick rotation of 
the time coordinate to the imaginary plane reflects a change of signature and 
thereby a topology change of the manifold describing a real tunneling geometry.  
Moreover, there have been recent suggestions that quantum gravity topology 
change is connected with Perelman's Ricci flow 
\cite{Frenkel:2020dic,Dzhunushaliev:2009jx}, 
while  in \cite{Hebecker:2018ofv} it was argued that a  topology change cannot 
be forbidden due to ``censorship'' theorems, since it is a well understood 
feature of string theory in 2D and 10D 
\cite{Tamvakis:1989aq,Bergshoeff:2004pg,Bergshoeff:2005zf,Hertog:2017owm}.
Finally, in \cite{Villani:2021aph,Duston_2020} the evolution of the topology   
was illustrated in loop quantum gravity with topspin network formalism.

Among others, topology changes may in principle affect the field equations that 
arise through the semiclassical variation procedure of gravitational actions. 
Although in the case of Einstein-Hilbert action this procedure
reproduces the standard field equations, one could 
investigate whether variation of the Gauss-Bonnet action on a topologically 
altered spacetime due to the formation of microscopic wormholes could lead to 
a non-trivial result. Interestingly enough, such an extended analysis 
induces extra terms in the field equations, which can be interpreted as an 
effective dark energy sector of topological origin.

This manuscript is organized as follows:  In Section \ref{Topology change in 
EQG} we briefly review the topology change in the framework of Euclidean 
quantum gravity induced by objects with distinct topology, and in 
Section \ref{ETVC}  we present the effective topological variation procedure. 
Then in section \ref{Semiclassical Einstein} we apply this procedure and we 
derive the semiclassical field equations for the cases of Einstein-Hilbert and 
Gauss-Bonnet actions, while in Section \ref{Results} we show the appearance of 
an effective dark energy induced by microscopic Gauss-Bonnet 
wormholes. Finally, Section \ref{Conclusion} is devoted to the conclusions

\section{\label{Topology change in EQG}Topology change in Euclidean Quantum 
Gravity}

In this section we discuss the effects of topology change in the framework of 
Euclidean quantum gravity (EQG). In EQG context, in order for the complex 
path integral to converge, the time dimension is Wick rotated $t\rightarrow 
i\tau$, thus the Lorentzian signature $(-+++)$ changes to Euclidean $(++++)$. In 
the complex  path integral there are saddle point solutions, which correspond 
to classical solutions, namely  instantons, with different topology from the 
background \cite{HAWKING1978349,Gibbons:1978ac} and therefore they mediate 
topology change \cite{Gibbons:2011dh}. These solutions can represent the 
creation of a pair of black holes or Euclidean wormholes under a strong field 
as in Schwinger process \cite{Gibbons:2011dh,Garattini:2000ge}. In the sum over 
history approach, Sorkin \cite{Sorkin:1997gi} has developed a calculus for 
topology change based on Morse theory, where the transition between two manifolds 
of distinct topology is being performed by a Morse function provided a 
cobordism exists between the manifolds \cite{Dowker:1997kc,Dowker:2002hm}. 
 
The topological structure of a manifold $M$ is  characterized by topological 
indices, and one of the most extensively studied is the Euler characteristic 
$\chi(M)$. 
In the modern language of differential forms and in the context of de Rham 
Comohology, the Euler characteristic $\chi$ is defined as the alternating sum of the 
Betti numbers of the manifold $M$ \cite{Nakahara:2003nw}
\begin{equation}\label{chi_Betti}
    \chi(M)=\sum_{p}(-1)^{p}B_{p},
\end{equation}
where the Betti numbers $B_{p}$ of a manifold are defined as the dimension of 
the $p^{th}$ de Rahm cohomology group \cite{Nakahara:2003nw}
\begin{equation}\label{Rham B to H}
    B_{p}=dim H^{p}(M).
\end{equation}
In the above expression the $p^{th}$ de Rahm cohomology group $H^{p}(M)$ is the 
set of all closed 
p-forms $Z^{p}(M)$ modulo the set of all exact p-forms $B^{p}(M)$
\begin{equation}\label{Rham form mod}
    H^{p}(M)=Z^{p}(M)/B^{p}(M),
\end{equation}
where a closed form satisfies $d\omega=0$ (where $\omega$ is 
a $p$-form and $d$ denotes the exterior derivative), and  an exact form 
satisfies  $\omega=dn$ (where $n$ is 
a  $p$-form). 

The Poincare Lemma states that a closed form  defined on a 
domain $V\subseteq M$ is also exact,   if the domain $V$ is 
contractible to a point. In the light of Poincare Lemma, de Rham Cohomology 
can be seen as a restriction on the global exactness of closed forms 
\cite{Nakahara:2003nw}. In summary, 
Betti numbers measure the global inexactness of closed forms as obstructions 
to contractibility to a point, originated from holes and discontinuities of the 
domain \cite{rodriguez2015signature}.

In Table \ref{tab:table1} we present the value of  Euler characteristics 
  for different spacetime manifolds.
\begin{table}[ht]
\begin{ruledtabular}
\begin{tabular}{lcdr}
\textrm{Spacetime}& $\!\!\!\!$\textrm{Euler characteristic} \\ 
 $\ $&   \protect$\chi$\\ 
\hline
 \textrm{Minkowski} & 0\\
 \textrm{Extreme Black Holes} & 0\\
 \textrm{Self-dual Taub-Newman-Unti-Tamburino  } & 1\\
 \textrm{Schwarchild and Kerr Black Holes} & 2\\
 \textrm{Nariai}\;$S_{2}\times S_{2}$ & 4\\
 \textrm{Euclidean Wormhole}\;$S_{1}\times S_{3}$ & 0 
\end{tabular}
\end{ruledtabular}
\caption{\label{tab:table1}%
Euler characteristics  for different spacetime manifolds as it has been 
calculated   in 
\cite{Gibbons:1978ac,Gibbons:1979xm,Liberati:1995jj,Gibbons:1994ff,Ma:2003uj}. 
By the product property for product manifolds
$ \chi(M_{1}\times M_{2})=\chi (M_{1}) \cdot \chi (M_{2})$,
one can easily verify that for a Nariai instanton $\chi_{Na}=2\cdot 2=4$ and 
similarly for a Euclidean wormhole $\chi_{EW}=0 \cdot 3=0$. }
\end{table}

In order to investigate the topology change, one can decompose a 
4D  manifold $M$ into a connected sum (symbolized by $\#$) of two 4D
manifolds $M_{1}$ and $M_{2}$, by gluing them together at the boundaries left by 
the removal of a four-ball. For connected sums,   
the Euler characteristic is given by \cite{Nakahara:2003nw}
\begin{equation}
    \chi\left(M_{1}\#M_{2}\right)=\chi(M_{1})+\chi(M_{2})-2.
\end{equation}
Then, following Gibbons \cite{Gibbons:1991tp,Gibbons:2011dh}, the formation of a 
Euclidean wormhole with topology $(S_{1}\times S_{3})$, namely
\begin{equation}
    M\rightarrow M\#(S_{1}\times S_{3}),
\end{equation}
decreases $\chi$ by $2$, thus $\delta\chi=-2$ , while the formation of a  Nariai 
instanton with topology $(S_{2}\times S_{2})$, namely
\begin{equation}
    M\rightarrow M\#(S_{2}\times S_{2}),
\end{equation}
increases $\chi$ by $2$ thus $\delta\chi=2$.

Since the formation of  gravitational instantons or wormholes change the Euler 
characteristic of the 4D spacetime, one must examine  its effect in a 
systematic way. For a 4D spacetime manifold equation 
 \eqref{chi_Betti} becomes  \cite{rodriguez2015signature}
\begin{equation}\label{chi rel Betti}
    \chi=b_{0}-b_{1}+b_{2}+b_{3},
\end{equation}
with   $b_{0}$   the number of 
connected components, $b_{1}$ the number of one-dimensional holes, $b_{2}$ the 
number of two-dimensional holes, and $b_{3}$   the number of three-dimensional 
holes. Hence, when the Euler characteristic changes, there is a change in the 
Betti numbers of spacetime \cite{Schulz:2018fun}, which corresponds to a change 
in the dimension of De Rham Cohomology group and therefore to a change in  the 
proportion of exact to closed forms. The induced non-triviality of De Rham 
Cohomology group indicates that a closed form exists which is no longer exact. 
Therefore, by the Poincare Lemma,  the induced inexactness implies that there 
is an area of the manifold,  a wormhole,  which is not contractible to a point.

In summary, the formation for a wormhole changes the topology and this 
property will be exploited in the following.

\section{\label{ETVC}The effective topological variation procedure}

Inspired by Wheeler's conceptualization  of spacetime foam 
\cite{wheeler1964relativity}, where quantum fluctuations of the metric are 
considered to cause fluctuations of the topology of the spacetime manifold, 
which was later developed in the context of Euclidean quantum gravity by 
Hawking, Gibbons, Sorkin and others 
\cite{HAWKING1978349,hawking1993euclidean,Sorkin:1985bh}, we are interested in 
investigating the behavior of the variation of  higher-order gravitational 
actions
under the assumption that the variation of the quantum 
field fluctuations  $\delta h$ causes a variation in the topology of the 
spacetime manifold $\delta\chi$.

\begin{figure}
    \centering
    \includegraphics{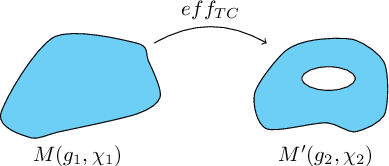}
    \caption{{\it{A three-dimensional  illustration of the effective topology change 
from a manifold of Euler characteristic \protect$\chi_{1}$ to a manifold of 
Euler characteristic \protect$\chi_{2}$}.}}
    \label{fig:effTC}
\end{figure}

We consider that the process of signature change of the spacetime 
manifold by the Wick rotation and the Euclidean quantum gravity path integral 
convergence contour deformation, which yields instaton solutions of different 
spacetime topology, can be encapsulated into an effective topology change  
operation ($eff_{TC}$), as illustrated in Fig.~\ref{fig:effTC}. Specifically, 
for manifolds with metric $g_i$ and  Euler characteristics $\chi_{i}$,  it can 
be encapsulated into the variation of the gravitational quantum field $\delta 
h$, namely
 \begin{align}\label{encaps}
    M&(g_{1},\chi_{1})\xrightarrow{eff_{TC}} M^{'}(g_{2}.\chi_{2}),\nonumber\\
    &eff_{TC}: \delta h \longrightarrow \delta\chi.
 \end{align}

  One could argue that EQG topology change is a 
mathematical artifact, a byproduct of the Wick rotation and of the integral 
contour  deformation. However, there have been many studies 
supporting that topology change occurs in other quantum gravity approaches 
too, for instance in string theory 
\cite{Tamvakis:1989aq,Bergshoeff:2004pg,Bergshoeff:2005zf,Hertog:2017owm} 
and loop quantum gravity \cite{Villani:2021aph,Duston_2020}. Therefore, it seems 
likely that topology change is a generic feature of quantum gravity 
\cite{Sorkin:1997gi,Hebecker:2018ofv}. An additional objection could be that 
processes  which change the manifold topology   may lead to 
the development of singularities, in the cases where they give rise to discrete 
 and/or non-differentiable submanifolds, which could  make the 
variational calculus ambiguous. However, in 3D one can introduce 
novel surgery techniques that successfully treat the singularities that develop 
in 3D manifolds during their Ricci flow  \cite{Perelman:2006up}, and thus one 
could in principle follow the same procedure in 4D manifolds. Nevertheless, we 
should comment here that since 4D topologies remain un-classifiable 
\cite{Geroch:1986iu}, the path integral approach is still not fully 
well-defined mathematically, and since topology change can emerge by 
the path integral approach \cite{Gibbons:2011dh},  it is interesting to 
explore the implications of  relation \eqref{encaps} as long as it may lead to 
interesting novel physical results. We will investigate a variational 
formulation of topology changes in a separate work. 

Let us examine the above consideration in more detail. A common technique in 
many approaches to quantum gravity is to split linearly the full metric $g$ into 
a background metric $\Tilde{g}$ and the quantum fluctuation field $h$ around it \cite{Knorr:2021slg}
\begin{equation}\label{linear split}
    g^{\mu\nu}=\Tilde{g}^{\mu\nu}+h^{\mu\nu}.
\end{equation}
According to Wheeler's argument \cite{Wheeler:1955zz,Wheeler:1957mu} and similar 
more recent ones \cite{Hossenfelder:2012jw}, the quantum fluctuations are 
scale-dependent as $\delta h\sim \frac{l_{p}}{l}$,  and they become large near the Planck scale, remaining always smaller than one as long as we consider the Planck scale as cut-off. However, from the viewpoint of asymptotic 
safety, one could argue that interactions could become weak at the Planck scale 
by an appropriate renormalization-group flow \cite{Hossenfelder:2012jw}. If 
one follows the first consideration, fluctuations may induce topology 
change, nevertheless  higher-order terms  in the expansion could be non negligible. 
If one follows the second consideration, then classical expansion techniques can 
be employed but ambiguities arise on the ability of fluctuations to induce
topology change. Since a solid theory of quantum gravity remains far from being 
complete, 
in this work
we assume that a compromise exists between 
the two extremes, suggesting that quantum fluctuations can produce topology 
change at small scales,
while being small enough in order for perturbation theory to hold. Consequently, this scaling approach facilitates a reduction in the significance of higher-order terms, thereby allowing the phenomena to be predominantly described by the first-order term.

As mentioned in \cite{Pawlowski:2020qer},  the quantum fluctuation field 
$h=g-\Tilde{g}$ of the linear split is not a metric and lacks a geometrical 
meaning, therefore among the other types of split,  it is the most suited for 
describing the fluctuation field that causes topology change. Consequently, we   
consider $\Tilde{g}_{\mu\nu}$ to be the dynamical background metric and we 
handle $h_{\mu\nu}$ as an effective ``matter'' field. 

We close this section by commenting on the background independence and the 
tadpole condition. In particular, in most treatments of quantum gravity there 
are strong arguments in favor of the independence of the quantum effective action 
from the background, since physical observables must be background 
independent and the action must be diffeomorphism invariant. By demanding 
background independence, one can introduce  split symmetry  
\cite{Pawlowski:2020qer}, given by all the transformations of the background 
metric and   fluctuation fields that preserve the full metric, namely
\begin{equation}
    g(\Tilde{g},h)\rightarrow g(\Tilde{g}+\delta \Tilde{g},h+\delta 
h)=g(\Tilde{g},h).
\end{equation}

In the quantization scheme of \cite{Becker:2021pwo}, background independence is 
guaranteed by the class of metrics that are self consistent. Self 
consistent metrics $ \Tilde{g}^{SC}$ are those that allow the effective 
field equations obtained from the effective action 
$\Gamma[h_{\mu\nu},\Tilde{g}]$ to admit the solution $h_{\mu\nu}=0$. Thus, if 
one incorporates background independence into the extremization condition of the 
effective action,  then one obtains the tadpole condition \cite{Becker:2021pwo}
\begin{equation}\label{tadpole}
    \frac{\delta}{\delta h_{\mu\nu}}\Gamma 
[h,\Tilde{g}]\bigg|_{h=0,\;\Tilde{g}=\Tilde{g}^{SC}}=0.
\end{equation}

\section{\label{Semiclassical Einstein}
Effective cosmological constant  of topological origin} 

We now have all the machinery to perform the variation of gravitational actions in 
cases where there are topology changes in the underlying spacetime manifold. 
Firstly, we will apply the procedure in the case of Einstein-Hilbert 
action, which leads to a trivial result, and then we will apply it in the case 
of the Gauss-Bonnet modified action, where we will see the surprising result of 
the appearance of an effective cosmological constant.

\subsection{Einstein-Hilbert 
action}\label{The semiclassical approach for the EH action} 

Let us start by  presenting the semiclassical approach described above in the 
case of Einstein-Hilbert action, as it has been demonstrated firstly by 
`t Hooft in 
\cite{tHooft:1974toh}. By performing a Wick rotation, the spacetime 
signature becomes $(+,+,+,+)$ and the action will be Euclideanized, i.e.
\begin{equation}
    S_{EH}=-\frac{1}{2\kappa^2}\int d^{4}x\sqrt{g}R, 
\end{equation}
with $\kappa^2$ the gravitational constant.
We  expand the Einstein-Hilbert action  around  the background field 
according to the metric split  (\ref{linear split}), in orders of the ``quantum 
field'', namely
\begin{equation}
    S_{EH}=S_{0}+S_{1}+S_{2}+\sum_{n=3}^{\infty}S_{n}.
\end{equation}
In order to calculate  each term we expand the inverse 
metric as
\begin{equation}\label{g expansion}
    g^{\mu\nu}=\Tilde{g}^{\mu\nu}-h^{\mu\nu}+h^{\mu}_{\lambda}h^{\lambda\nu}+\mathcal{O}(h^{3}),
\end{equation}
and then by employing the property $\log \det A= \text{tr} \log A$ and 
performing logarithmic and exponential expansions, we express the determinant of 
the metric in terms of powers of $h$ as
\begin{equation}\label{det g expansion}
    \sqrt{g}=\sqrt{\Tilde{g}}\left[ 1+\frac{1}{2} \Tilde{g}_{\mu\nu} 
h^{\mu\nu}-\frac{1}{4}h^{\mu\nu}h_{\mu\nu}+\frac{1}{8}(h^{\mu}_{\mu})^{2}
+\mathcal{O}(h^{3}) \right], 
\end{equation}
where the uppering and  lowering of the  quantum field  indices  are 
performed using  the background metric, namely
$h=h^{\mu}_{\mu}=\Tilde{g}^{\mu\nu}h_{\mu\nu}$. Since the Ricci scalar is  
$R=g^{\mu\nu}R_{\mu\nu}$,   by employing  (\ref{g expansion}) and 
 (\ref{det g expansion})  the Ricci tensor and   Ricci scalar can be 
expanded in the same manner. After some algebra  and 
neglecting the total derivatives, the first three terms of the  Einstein-Hilbert 
expansion are expressed as 
\begin{align}\label{EH expansion}
    S_{0}&=-\frac{1}{2\kappa^2}\int d^{4}x \sqrt{\Tilde{g}} \Tilde{R}\nonumber\\
    S_{1}&=\frac{1}{2\kappa^2}\int d^{4}x \sqrt{\Tilde{g}}\left( \Tilde{R}_{\mu\nu}-\frac{1}{2}\Tilde{g}_{\mu\nu}\Tilde{R}\right)h^{\mu\nu}\nonumber\\
    S_{2}&=-\frac{1}{2\kappa^2}\int d^{4}x \sqrt{\Tilde{g}}\biggl\{ \frac{1}{4}h^{\mu\nu} \nabla^{2} h_{\mu\nu}-\frac{1}{8}h \nabla^{2} h\nonumber\\
    &+\frac{1}{2}\left( \nabla^{\nu}h_{\nu\mu}-\frac{1}{2}\nabla_{\mu}h\right)^{2}+\frac{1}{2}h^{\mu\lambda}h^{\nu\sigma}\Tilde{R}_{\mu\lambda\nu\sigma}\nonumber\\
    &+\frac{1}{2}\left(h^{\mu\lambda}h^{\nu}_{\lambda}-h h^{\mu\nu}\right)\Tilde{R}_{\mu\nu}+\frac{1}{8}\left(h^{2}-2h^{\mu\nu}h_{\mu\nu}\right)\Tilde{R}\biggr\}.
\end{align}

In summary, the effective action up to one-loop approximation will be
\begin{equation}\label{EH effective}
    \Gamma=S_{EH}+\Gamma_{1L}+\mathcal{O}(2-loop),
\end{equation}
where the quadratic terms of the quantum field $h$  are absorbed in the 
one-loop part
\begin{equation}
    \Gamma_{1L}=\Gamma_{GF}+\Gamma_{Fgh},
\end{equation}
with $\Gamma_{GF}$ and $\Gamma_{Fgh}$ corresponding to the effective action for 
the gauge fixing and ghost terms respectively \cite{Becker:2021pwo}.
One can then vary  the Einstein-Hilbert action   due to quantum fluctuations of 
the field 
$h^{\mu\nu}\rightarrow h^{\mu\nu}+\delta h^{\mu\nu}$,  i.e.  calculate 
$    
\delta_{h} S_{EH}$. Imposing the tadpole condition  
(\ref{tadpole}) for the effective action  \eqref{EH effective} and taking into 
account \eqref{EH expansion}, one finally retrieves the Einstein equations 
for the classical background as \cite{Becker:2021pwo,Pagani:2019vfm}
\begin{equation}
    \Tilde{R}_{\mu\nu}-\frac{1}{2}\Tilde{g}_{\mu\nu}\Tilde{R}=\kappa^2 
T_{\mu\nu},
\end{equation}
where the stress tensor originates from the one-loop part of the effective 
action, containing   matter as correction terms in the right-hand-side, i.e.
\begin{equation}\label{eff stress tensor}
    T^{\mu\nu}=-\frac{2}{\sqrt{g}}\frac{\delta}{\delta 
h^{\mu\nu}}\Gamma_{1L}\big|_{h=0}.
\end{equation} 
Actually, this was expected, since the Einstein-Hilbert action term is the 
Euler density in two-dimensions, and thus 
its variation due to variations of the quantum field $h$ in 4D will be the  
standard one \cite{Padmanabhan:2013xyr}.

\subsection{ Gauss-Bonnet action}

Let us now perform the above procedure in the case of the Gauss-Bonnet action.
The Gauss-Bonnet (GB) curvature polynomial $\mathcal{G}$ is defined as
\begin{equation}
\mathcal{G}=R^{2}-4R_{\mu\nu}R^{\mu\nu}+R_{\mu\nu\rho\sigma} 
R^{\mu\nu\rho\sigma},
\end{equation}
 and it is known that in four dimensions such a term is a topological 
invariant. In order to see this in the context of the present manuscript, we 
recall that the Chern-Gauss-Bonnet theorem \cite{Chern:1945} states that for 
the case  of a 
compact orientable manifold $M$ with boundary $\partial M$ of  dimension $D=4$, 
the Euler characteristic is
\begin{equation}\label{Chern Gauss Bonnet}    
\chi(M)=\frac{1}{32\pi^{2}}\int_{M}d^{4}x\sqrt{g}\;\mathcal{G}+\int_{\partial 
M}Q,
\end{equation}
with $Q$ an appropriate correction form 
integrated on the boundary $\partial M$ \cite{Eguchi:1980jx}. 
The essence of the theorem is that despite any local 
deformation of the manifold, its total curvature, as expressed by the integral 
of the GB curvature polynomial, depends only on the topology of the manifold. 
Consequently, for a manifold of fixed topology, $\chi$ is considered a 
topological invariant under smooth variations of the metric 
\cite{Fernandes:2022zrq}. 
 
We can now perform the steps of the previous subsections in the case of the 
Gauss-Bonnet action. Its Euclideanized form is 
 \begin{equation}
      S_{GB}=-\frac{\alpha}{2\kappa^2}\int d^{4}x \sqrt{g}\left(R^{2}-4R^{\mu\nu}R_{\mu\nu}+R^{\mu\nu\rho\sigma}R_{\mu\nu\rho\sigma}\right),
\end{equation}    
where $\alpha$ is the coupling parameter.
The effective action under the one-loop approximation will be
\begin{equation}\label{effective Gamma}
    \Gamma= S_{GB}+
    \Gamma_{1L}+
    \mathcal{O}_{(2-Loop)}.
\end{equation}
Varying the GB action with respect to quantum fluctuations of the field 
$h^{\mu\nu}\rightarrow h^{\mu\nu}+\delta h^{\mu\nu}$   according to the 
topological variational procedure and applying  the  4D Chern-Gauss-Bonnet  
theorem \eqref{Chern Gauss Bonnet} without 
a boundary,   we obtain
\begin{align}
    &\!\!
\delta_{h} S_{GB}\nonumber\\
    &=\delta_{h}\left[-\frac{\alpha}{2\kappa^2}\int_{M} 
d^{4}x\sqrt{g}\left(R^{2}-2R_{\mu\nu}R^{\mu\nu}+R_{\mu\nu\rho\sigma}R^{
\mu\nu\rho\sigma}\right)\right]\nonumber\\
 &  =-32\pi^{2}\frac{\alpha}{2\kappa^2}\frac{\delta\chi}{\delta 
h^{\mu\nu}}\delta h^{\mu\nu}.
\end{align}
Implementing the substitution 
$\delta\chi\rightarrow \partial\chi$ and applying the chain rule, we find
\begin{align}\label{dSGB A}
    &
    \!\!\!\!\!\!\!\!\!\!\!\!\!\!\!\!\!
    \delta_{h}S_{GB}= -16\pi^{2}\frac{\alpha}{2\kappa^2}\frac{\partial \chi}{\partial 
V}\frac{\delta V}{\delta h^{\mu\nu}}\delta 
h^{\mu\nu}
\nonumber\\
&=-16\pi^{2}\frac{\alpha}{2\kappa^2}\frac{\partial \chi}{\partial V} 
\delta_{h}\left(\int_{M}d^{4}x \sqrt{g}\right)\nonumber\\
    &=-16\pi^{2}\frac{\alpha}{2\kappa^2}\frac{\partial \chi}{\partial V}\int_{M}d^{4}x 
\frac{\delta\sqrt{g}}{\delta h^{\mu\nu}}\delta h^{\mu\nu},
\end{align} 
with $V$ the manifold volume.
If one implements the expansion of the metric determinant \eqref{det g 
expansion}, then the functional derivative into the integral becomes
$ \frac{\delta\sqrt{g}}{\delta h^{\mu\nu}} 
= \frac{1}{2}\sqrt{\Tilde{g}}\Tilde{g}_{\mu\nu}+\mathcal{O}(h),
$ so after inserting it into \eqref{dSGB A} we finally acquire 
\begin{equation}
\delta_{h}S_{GB}=-16\pi^{2}\frac{\alpha}{2\kappa^2}\frac{\partial\chi}{\partial 
V}\int_{M} d^{4}x\sqrt{-\Tilde{g}}\Tilde{g}_{\mu\nu}\delta h^{\mu\nu}.
\end{equation}
Finally, we make the reasonable  approximation that for an infinitesimal 
integration volume the topology change per volume $\frac{\partial \chi}{\partial 
V}$ remains constant and thus it can enter inside the integral,  in which case  
   the topological variation of the Gauss-Bonnet term is expressed as
\begin{equation}\label{fderivative GB}
    \frac{1}{\sqrt{\Tilde{g}}}\frac{\delta S_{GB}}{\delta h^{\mu\nu}}=- 
16\pi^{2}\frac{\alpha}{2\kappa^2}\frac{\partial\chi}{\partial 
V}\Tilde{g}_{\mu\nu}+\mathcal{O}(h).
\end{equation}
Interestingly enough,  the 
variation of the Gauss-Bonnet term on a manifold that has topology changes due 
to the formation of wormholes is not zero.

\subsection{Einstein-Gauss-Bonnet 
action}

Let us now consider the full case of general relativity plus a Gauss-Bonnet 
correction, namely 
\begin{equation}
 S_{tot}=S_{EH}+S_{GB}.
\end{equation}
As we analyzed above, although the 
Einstein-Hilbert term gives the standard classical field equations, the 
Gauss-Bonnet term leads to a non-trivial semi-classical result.  
In particular, the effective action under the one-loop approximation will be
\begin{equation}\label{effective Gamma GB}
    \Gamma= S_{EH}+S_{GB}+
    \Gamma_{1L}+
    \mathcal{O}_{(2-Loop)}.
\end{equation}
 Calculating the fuctional derivative of the effective 
action by employing Eq.~\eqref{EH expansion} and 
Eq.~\eqref{fderivative GB}, we finally obtain
\begin{align}
    &\frac{1}{\sqrt{\Tilde{g}}}\frac{\delta\Gamma}{\delta h^{\mu\nu}}=
    \frac{1}{\sqrt{\Tilde{g}}}\frac{\delta S_{EH}}{\delta 
h^{\mu\nu}}+\frac{1}{\sqrt{\Tilde{g}}}\frac{\delta S_{GB}}{\delta 
h^{\mu\nu}}+\frac{1}{\sqrt{\Tilde{g}}}\frac{\delta\Gamma_{1L}}{\delta 
h^{\mu\nu}}\nonumber\\    
&=\frac{1}{2\kappa^{2}}\biggl\{\!\Tilde{R}_{\mu\nu}\!-\!\frac{1}{2}\Tilde{g}_{ 
\mu\nu } \Tilde{R}\!-\!16\pi^{2} \alpha 
\frac{\partial\chi}{\partial 
V}\Tilde{g}_{\mu\nu}\!+\!\frac{1}{\sqrt{\Tilde{g}}}\frac{\delta\Gamma_{1L}}{
\delta 
h^{\mu\nu}}\!+\!\mathcal{O}(h)\!\biggl\}.
\end{align}
Hence, imposing the tadpole condition  \eqref{tadpole}  that removes the 
terms $\mathcal{O}(h)$, we find the semi-classical field equations
\begin{equation}
\Tilde{R}_{\mu\nu}-\frac{1}{2}\Tilde{g}_{\mu\nu}\Tilde{R}+\Tilde{g}_{\mu\nu}
\Lambda_{eff}=\kappa^2 T_{\mu\nu},
\end{equation}
where the stress tensor is given by \eqref{eff stress tensor}, and where we 
have defined 
\begin{equation}\label{Leff}
    \Lambda_{eff}\equiv-16\pi^{2}\alpha \frac{\partial \chi}{\partial V}.
\end{equation}
As we observe, we have obtained an effective cosmological constant term of 
topological origin, induced by the Gauss-Bonnet correction term due to the 
topology change that  microscopic wormholes brought about. This 
is the main result of the present work.

\section{Dark energy from  microscopic Gauss-Bonnet   wormholes}\label{Results}

As we showed in the previous sections, the variation of gravitational 
actions in cases where one has topology changes in the underlying spacetime 
manifold, may lead to extra terms in the field equations. Although in the case 
of Einstein-Hilbert action one does not obtain any non-trivial result, 
incorporating the Gauss-Bonnet correction on such topologically-changed 
manifolds gives rise to an effective cosmological constant, even in 4D where 
the GB term is known to have no effect at the classical level.  
As we discussed, according to the literature such topological changes can 
typically arise when objects like instatons or wormholes are formed at the 
microscopic level.
 
In particular, the effective cosmological constant is just  the term 
$\frac{\partial\chi}{\partial V}$, which can be interpreted as the density of 
the non-trivial microscopic objects per four-volume 
$\rho_{obj}=\frac{N_{obj}}{V}$, since these objects induce the topology change 
(for instance $\delta\chi=-2$ 
corresponds  to the formation of a Euclidean wormhole, while 
$\delta\chi=2$ to the formation of a Nariai instanton).
Hence, according to \eqref{Leff}, the 
effective cosmological constant equals the density $\rho_{w}$ of  
$N_{w}$ topology 
changing wormholes  per four-volume, i.e.
\begin{equation}\label{wormhole density}
    \Lambda_{eff}=-16\pi^{2}\alpha \rho_{w}=-16\pi^{2}\alpha\frac{N_{w}}{V},
\end{equation}
namely  it  depends on the GB coupling parameter 
$\alpha$ and on the wormhole density.

Concerning the value of  GB coupling, there is a consensus   that since the GB 
term   appears  in the low-energy limit of an effective action 
\cite{Nojiri:2005jg},      $\alpha$    is related to 
the inverse of the string tension 
\cite{Kanti:1995vq,Ong:2022mmm,Zwiebach:1985uq} 
$\alpha\sim(1/ \sqrt{a^{'}})$ or equivalently to the square of the string scale 
$\alpha \sim l^{2}_{s}$ \cite{Bamba:2014zoa,Padmanabhan:2013xyr}. Since the 
string scale $l_{s}$ cannot be far from the Planck scale in four dimensions 
$l_{s}\sim l_{p}$ \cite{Berenstein:2014wva,Gubser:2003vk}, one first estimation 
could be $\alpha=l^{2}_{p}$. In such a case, if we identify the effective 
cosmological constant  $  \Lambda_{eff}$ of (\ref{wormhole density}) with the observed cosmological constant 
$\Lambda_{obs}=10^{-52}m^{-2}$, we need a microscopic wormhole density of 
 $\rho_{w}=10^{16}$ wormholes per cubic meter per second, which is quite 
reasonable according to Hawking and Schulz estimations for the spacetime  foam 
\cite{HAWKING1978349,Anderson:1986ww,Schulz:2018fun}.
On the other hand, since the upper  bound for the wormhole density is 
one wormhole per Planck volume, namely 
$\rho_{Mw}=\frac{1}{l^{4}_{p}}\sim 10^{140}$, according to 
 \eqref{wormhole density} the upper bound of $\Lambda_{eff}$ is
$\Lambda_{M}\sim 10^{72} m^{-2}$, or approximately $10^{124}$ larger than 
$\Lambda_{obs}$. 

We stress here that the wormhole density in a dynamical 
spacetime is not expected to be constant, therefore the obtained effective 
cosmological constant also acquires a dynamical nature, i.e. it corresponds to 
an effective dark energy sector.
 
Note that in different frameworks there have also been  approaches where the 
cosmological constant is driven by spacetime wormholes, but they typically have
 $\Lambda\rightarrow0$ at late times. For instance, in 
\cite{HAWKING1978349} Hawking considers space time foam as a gas of instatons 
of different topology and in the Euclidean quantum gravity one-loop 
approximation he obtains a negative cosmological constant   
$\Lambda_{s}\sim\alpha\frac{\chi}{V}$, which although having been extracted in a 
totally different framework, it resembles our result \eqref{Leff}. However, 
Hawking's calculations are based on the trace anomaly expressed by the 
invariant GB term, and for that reason $\chi$ appears constant (refinements of 
Hawkings spacetime foam model were presented in 
\cite{Christensen:1979iy,Schulz:2018fun}). In \cite{Coleman:1988tj}, Coleman 
proposed a mechanism where wormholes and topological fluctuations of space 
time induce a distribution of the values of nature's constants, which smear 
$\Lambda$  distribution to peak at zero. In \cite{Grinstein:1988eb,Ellis:1989vk} 
it was claimed that the behavior of the fundamental coupling constants in 
Coleman's scenario was controlled by the trace anomaly and a similar 
proposition was the Giddings-Strominger wormhole solution 
where a wormhole is coupled to an instanton \cite{Giddings:1987cg}. 
Additionally,
in \cite{Garattini:2000ge}  a semiclassical model of spacetime foam was 
proposed, in which Casimir-like quantum fluctuations give rise to an 
arbitrary number of wormholes, as pairs of  black-hole and anti-black-hole, which 
drive the induced cosmological constant to zero as they grow. 

Nevertheless, in 
our approach microscopic wormholes lead to an effective dark energy sector not 
directly, but due to the topology change they induce on the manifold, which in 
turn affects the variation of the GB term. That is why it can have an arbitrary 
dynamical behavior. Additionally, apart from its effects at late-time 
cosmology, such a dynamical effective sector could play a role in the early 
universe too, potentially driving inflation. Finally, note that since Nariai instatons 
correspond to a negative component while Euclidean wormholes to a positive 
one, one could have richer cosmological behavior as well.
 
\section{Conclusions}
\label{Conclusion} 
 
It is known that the appearance of microscopic objects, such as instatons and 
wormholes, at the spacetime-foam level in Euclidean quantum gravity approaches,
 leads to spacetime topology changes, which in principle may affect the field 
equations that arise through the variational procedure of gravitational actions. 
Although in the case of Einstein-Hilbert action the presence of microscopic 
wormholes does not lead to any non-trivial result, when the Gauss-Bonnet term 
is added in the action the above procedure induces an effective cosmological 
term that depends on the Gauss-Bonnet coupling and the wormhole density. Since 
the later in a dynamical spacetime is in general time-dependent, one results 
with an effective dark energy sector of topological origin.

In particular, the appearance of objects with distinct topology and 
thus with different Euler characteristics,
leads to a change of the topological character of the spacetime manifold. This 
process can be  encapsulated into an effective approach in which  
the variation of the quantum fluctuations induces a variation in the Euler 
characteristic, constituting the effective topological variation procedure.
By employing the semiclassical one-loop approach on the linear  split of the 
metric and, additionally, incorporating the background independence through the 
tadpole condition, we showed that the variation of the Gauss-Bonnet term in the 
Lagrangian gives rise to a non-trivial term in the field equations.
The obtained effective cosmological constant can coincide with the observed 
value $10^{-52}m^{-2}$ for densities of the order of  $10^{16}$ microscopic 
wormholes per cubic meter per second, which is quite reasonable according to 
estimations. 

It would be interesting to consider scenarios of time-dependent wormhole 
density and investigate the behavior of the resulting dynamical dark energy 
sector, including the  confrontation with observational data from Supernovae 
Type I (SNIa), Baryonic Acoustic Oscillations (BAO), and
Cosmic Microwave Background (CMB) observations, as well as with direct
Hubble constant measurements through cosmic chronometers (CC). Additionally, 
one could examine the matter perturbation evolution in such a dynamical 
scenario. Moreover, one could apply the same considerations at early times and 
examine the possibility of a successful inflation realization. At the more 
theoretical level, one could investigate the effective topological variation 
procedure going beyond the linear expansion level, as well as examine the  
effect of a topologically dynamical GB term in the trace anomaly behavior, in 
heterotic strings renormalizability, and in M-theory's $\beta$-function. All 
these studies extend beyond the scope of this manuscript and will be performed 
in future projects.

\subsection*{Acknowledgments} 
The authors would like to acknowledge the 
contribution of the COST Action 
CA21136 ``Addressing observational tensions in cosmology with systematics and 
fundamental physics (CosmoVerse)''. CT acknowledges   financial support from the 
A.G. Leventis Foundation.

\bibliography{DEGBW_prd_pub}

\end{document}